\documentclass[12pt,preprint]{aastex}

\slugcomment{}

\shorttitle{Resolved Dust Emission in a z=3.65 Quasar}
\shortauthors{Clements et al.}

\begin{document}

\title{Resolved Dust Emission in a Quasar at z=3.65}

\author{D.L. Clements}
\affil{Physics Department, Imperial College, Prince Consort Road, London SW7 2AZ, UK}

\author{G. Petitpas}
\affil{Submillimetre Array, Hilo, Hawaii, USA}

\author{D. Farrah}
\affil{Astronomy Centre, University of Sussex, Brighton, BN1 9QH, United Kingdom}

\author{E. Hatziminaoglou}
\affil{European Southern Observatory, Garching, Munich, Germany \& IAC,Tenerife, Spain}

\author{T. Babbedge, M. Rowan-Robinson}
\affil{Physics Department, Imperial College, Prince Consort Road, London SW7 2AZ, UK}

\author{I. P\'erez-Fournon, Antonio Hern\'an-Caballero, Nieves Castro-Rodr\'\i guez}
\affil{IAC, Tenerife, Spain}

\author{C. Lonsdale}
\affil{IPAC, Caltech, Passadena, CA, US}

\author{J. Surace}
\affil{Spitzer Science Centre, Caltech, Pasadena, CA, USA}

\author{A. Franceschini}
\affil{Universita di Padova, Padova, Italy}

\author{B.J. Wilkes}
\affil{Harvard-Smithsonian Centre for Astrophysics, Cambridge, MASS, USA}

\and

\author{H. Smith}
\affil{UCSD, San Diego, CA, USA}

\begin{abstract}
We present submillimetre observations of the z=3.653 quasar SDSS160705+533558 together with data
in the optical and infrared. The object is unusually bright in the far-IR and submm with an IR luminosity of $\sim10^{14} L_{\odot}$. We ascribe this luminosity to a combination of AGN and starburst emission, with the starburst forming stars at a rate of a few thousand solar masses per year.  Submillimetre Array (SMA) imaging observations with a resolution $\sim$1" show that the submm (850$\mu$m) emission is extended on scales of 10---35kpc and is offset from the optical position by $\sim$10 kpc. This morphology is dissimilar to that found in submm galaxies, which are generally un- or marginally resolved on arcsecond scales, or submm-luminous AGN where the AGN lies at the peak of the submm or molecular emission. The simplest explanation is that the object is in the early stages of a merger between a gas rich galaxy, which hosts the starburst, and a gas-poor AGN-host galaxy, which is responsible for the quasar emission. It is also possible that jet induced star formation might contribute to the unusual morphology.
\end{abstract}


\section{Introduction}
Understanding the rest-frame far-IR properties of high redshift galaxies and quasars has become a major field, inspired by the detection of the Cosmic Infrared Background (CIB) Puget et al., 1996; Fixsen et al., 1998), and by the unexpectedly large number of distant submm sources in SCUBA surveys (eg. Hughes et al., 1998; Smail, Ivison \& Blain, 1997; Eales et al., 2000; Coppin et al., 2006). Much of the far-IR emission in these objects appears to be produced by dust obscured starbursts (Alexander et al., 2005; Clements et al., 2008; Dye et al., 2008) similar to local ultraluminous Infrared galaxies (ULIRGs). While starbursts dominate the bolometric output of these systems they may still contain AGN. SED analysis of local ULIRGs (eg. Farrah et al., 2003) reveals that at least half contain an AGN, a figure confirmed by mid-IR spectral studies (Armus et al 2006, Farrah et al., 2007, Desai et al 2007). Meanwhile, correlations between galaxy bulge and supermassive black hole masses suggest that star formation and AGN fueling are intimately linked (Haring \& Rix, 2004). Studies of sources containing both starbursts and AGN probe this link. A number of far-IR luminous high z AGN are already known (eg. Priddey et al., 2003), some of which have associated submm sources. Separations between submm and AGN components in these sources cover the full range of scales on which submm emission can be resolved, ranging from 400kpc (Stevens et al., 2004) for sources observed by SCUBA (resolution 15") to 4" for the famous resolved submm-luminous AGN BR1202-0725 at z=4.69 (Omont et al., 1996). 

As part of efforts to understand the links between star formation and AGN activity at high redshift, we have been searching for $z>3$ objects that show evidence for both an AGN, and for a far-IR excess, and observing them at sub-mm wavelengths. One such object is {\objectname SDSS160705+533558}, a Quasar found in the Sloan Digital Sky Survey (SDSS). with  a spectroscopic redshift of z=3.653. It was observed during the Spitzer SWIRE survey (Lonsdale et al., 2004) and detected at 70$\mu$m and all shorter Spitzer bands (Hatziminaoglou et al., 2005) indicating strong rest-frame mid-IR emission. The source contains a weak CIV broad absorption line (BAL) system (Trump et al., 2006), a property possibly associated with enhanced far-IR emission (Priddey et al., 2007). Consideration of the broad-band spectral energy distribution (SED) using a template fitting approach led Hatziminaoglou et al. to conclude that it is hyperluminous ($L>10^{13} L_{\odot}$) in the mid-IR, and suggested the source might have extreme luminosity at longer wavelengths. 

\section{Observations}
{\objectname SDSS160705+533558} was initially observed with both the SCUBA bolometer array at the JCMT, observing at 850$\mu$m and 450$\mu$m, and the MAMBO array at the IRAM 30m, observing at 1100$\mu$m. Observations at the JCMT were conducted on 18 March 2005 under Grade 1 conditions. The data were reduced using the SURF system with Mars as a primary calibrator and OH231.8 as a secondary calibrator. The source was detected at 850$\mu$m at 10$\sigma$ confidence. An excess flux was obtained at 450$\mu$m (2.5$\sigma$, 220 $\pm$ 90 mJy once calibration errors are included)  though for the rest of this paper this is treated as a 3$\sigma$ upper limit of 490mJy. IRAM 30m  observations were carried out during the Winter 2004/2005 pool observing season using the 117 element version of the MAMBO array (Kreysa et al. 1998) operating at a wavelength of 1.2mm (250 GHz) using standard on-off photometry observing mode in good conditions. Absolute flux calibration was established by observations of Mars and Uranus, resulting in a flux calibration uncertainty of about 20\%.The data were reduced with standard procedures in the MOPSIC package developed by R. Zylka. More details of these observations are given in Perez-Fournon et al. (in prep.). The fluxes are listed in Table 1. 

Given the bright flux at 850$\mu$m, we observed {\objectname SDSS160705+533558} with the Sub-millimeter Array (Ho et al. (2004)). The source was observed using extended configuration on 04 June 2006, and in compact configurations on 02 and 23 March 2007 and 04 June 2007. Observations
on 23 March 2007 and 04 June 2007 used all 8 SMA antennas, the rest of the nights had only 7 available. The resulting image created by combining all these data has $uv$-coverage from 6.5 k$\lambda$ to
218 k$\lambda$ at the observed LO frequency of 340 GHz. Observations were made in dry conditions, with $\tau_{225}$ ranging from 0.03 to 0.08. For all observations, the full 4 GHz bandwidth (2 GHz in each sideband separated by 10 GHz) was combined to achieve a total rms noise of 1.3 mJy in the final map. The SMA data were calibrated using the MIR software package developed at Caltech and modified for the SMA. Gain calibration was performed using the nearby quasars 1642+689, 1419+543, and 3C345. Absolute flux calibration was performed using realtime measurements of the system temperatures with observations of Titan and Neptune to set the scale, and bandpass calibration was done using 1921-293 or 3C273. Astrometric accuracy was determined by comparing the positions of the calibration objects 3C345 and 1642+689, using 3C345 as the reference and VLA astrometry to provide the true position. It was found to be accurate at the 0.2" level. The data were imaged using Miriad (Sault et al., 1995) and further analysis carried out using the KARMA software package.

\section{Results}

\subsection{Resolved Submm Emission}
Figure 1 shows the submm emission in {\objectname SDSS160705+533558}, derived from the SMA observations, overlaid on the SDSS optical image. The rest-frame far-IR emission is extended, consisting of a marginally resolved clump to the N and an arm extending to the S. There is also a possible companion source detected at the 3$\sigma$ level 2.2" (14.7kpc) to the W. The peak of the far-IR emission is offset from the quasar position by $\sim$1.5" to the N ($\sim$ 10kpc), with the emission arm extending roughly through the quasar position to the south. The overall length of the source, including the extended northern blob and the emission to the south, is 5.4", or 36kpc. The bright northern blob has a major axis size of $\sim$2.4", corresponding to a size of 16kpc. Comparison of the flux profile of this blob with that of the 1.33"x0.85" beam shows that any unresolved component contributes at most 60\% of its flux. The overall flux recovered from the interferometer at 850$\mu$m is $\sim$12mJy, which is significantly lower than the SCUBA flux, suggesting we have resolved out emission on scales $>$4". 

The spatial separation between the quasar and the peak of the submm emission, and the extended nature of the latter component, strongly suggest a physical distinction between the processes powering the optical emission and the far-IR. Furthermore, the IRAC and MIPS 24$\mu$m emission is centered at the position of the optical quasar, suggesting that emission in these bands is more strongly associated with the quasar than with the longer wavelength submm emission. The 70$\mu$m MIPS astrometry is not sufficiently accurate for a similar conclusion to be drawn. The optical/mid-IR is clearly associated with the AGN. This cannot be true for the extended submm emission for which the most likely source is star formation of some kind. 

\subsection{Spectral Energy Distribution}

Analysis of spectral energy distributions (SEDs) can provide constrains on the contributions of various components to the luminosity of galaxies. For the current source, where the imaging observations  indicate distinct contributions from both an AGN and star formation, this is especially important. We obtain constraints on the total IR (1-1000$\mu$m) luminosity and the contributions from the AGN and starburst by fitting the IR-to-submm fluxes with model SEDs from a library of SEDs for starburst (Efstathiou et al, 2000) and AGN (Rowan-Robinson et al., 1995) components. We use the methods described in Farrah et al. (2003) to determine the range of total, starburst and AGN luminosities consistent with the data, and the star formation rate is derived from the starburst luminosity. The resulting best SED fit is shown in Figure 2, with the range of parameters giving adequate fits summarized in Table 1. We also apply the independent SED fitting methods of Hatzminaoglou et al. (2008) to our source and these results are also shown in Figure 2. This approach fits the optical and mid-ir components of the SED using the grid of torus models from Fritz et al., 2006. Any excess that remains after this is fitted by an empirical starburst model selected from the following templates: Arp220, M82, NGC 1482, NGC 4102,  NGC 5253,  NGC 7714.  Neither of the fits is perfect and there are significant detailed differences between the two results. However, both independent approaches reach the same broad conclusions about {\objectname SDSS160705+533558}: (i) an AGN dominates its bolometric output as well as the rest-frame mid-IR (this matches earlier conclusions in Hatziminaoglou et al. (2008) based purely on {\em Spitzer} fluxes and our own conclusions based on the IRAC and MIPS 24$\mu$m astrometry); (ii) however, a massive obscured starburst is necessary to explain the strong rest-frame far-IR/submm emission, with a star formation rate $>$1000M$_{\odot} yr^{-1}$. Our source thus appears to be a combination of AGN and starburst, with both components being very luminous in the infrared. The derived total infrared luminosity, $\sim 10^{14}L_{\odot}$, confirms that {\objectname SDSS160705+533558} is an HLIRG (Rowan-Robinson, 2000) and that it is among the most luminous far-IR objects known. 

\section{Discussion}
The discovery of extended rest-frame far-IR emission, on scales from 16 to 36kpc, offset from the optical quasar is unexpected. The majority of observations of distant submm galaxies (SMGs) to date have revealed small scale or unresolved emission at these wavelengths (eg. Younger et al. (2008a,b), Tacconi et al., (2006)) consistent with the 'Maximum Starburst' model for galactic bulge formation described by Elmegreen (1999). If we scale the source sizes measured in Tacconi et al. (2006) to that expected for the maximum SFR derived for {\objectname SDSS160705+533558} we would expect a source 3.4kpc in size rather than the 16kpc found for the brightest emission region or the 36kpc found for the overall structure. Our source is thus inconsistent with this `Maximum Starburst' model, suggesting that this is not a complete model for the high-redshift submm population.

Some submm-luminous high-redshift AGN are extended in the submm.  BRI 1202-0725, one of the few high-redshift submm-AGN with a higher luminosity than {\objectname SDSS160705+533558}, consists of a pair of sources separated by 4", the brightest of which has a size of $\sim$3"x0.7" (Omont et al., 1996).  Other resolved far-IR luminous quasars include PSS J2322+1944 (Carilli et al., 2002a) at z=4.12, whose CO emission is found to be extended on a scale of 2", BRI 1335-0417 at z=4.4, which is resolved into two components separated by $\sim$1" (Carilli et al., 2002b), and 4c41.17 which contains two CO components 1.8" apart (De Breuck et al., 2005). The lensed quasar APM 08279+5255, in contrast, is unresolved by a 0.7"x0.65" beam (Krips et al., 2007). However, unlike the current source all these objects have their AGN closely associated with one of the far-IR or CO peaks. We do not yet have CO data for our source, but we do not expect the CO distribution to be radically different from the 850$\mu$m continuum emission, so once again {\objectname SDSS160705+533558} is unusual.

One possible equivalent object is 4C60.07, a z=3.8 submm bright (23.8 $\pm$3.5mJy) radio galaxy studied by Ivison et al (2008), whose far-IR luminosity is less than that of {\objectname SDSS160705+533558}, where the submm components lie 10-30kpc on the plane of the sky away from the radio core. This source is thought to be a young merger between two high mass objects. The host of the offset radio nucleus is thought to have exhausted or expelled its molecular gas while the other is gas rich and responsible for much of the submm emission. The remaining submm emission comes from gas rich tidally stripped material falling towards the radio core. 4C60.07 and {\objectname SDSS160705+533558} are far from identical and more information regarding the molecular gas and underlying old stellar population in our source are needed before we can draw a direct comparison, but we speculate that the extended submm tail in our source is a tidal structure similar to that suggested for 4C60.07. The object might thus be in the early stages of a merger between a galaxy rich in molecular material and a dust poor galaxy which hosts a supermassive black hole. The interaction has triggered star formation in the dust rich system and quasar activity through fueling the supermassive black hole. The differences between this system and more conventional SMGs would then be a result of the early stage of the merger. An alternative is the possibility that some or all of the star formation in this system is triggered by interaction between jets from the AGN and molecular material (eg. Klamer et al., 2004). Observations at radio wavelengths in search of these jets would be needed to test this idea.

\section{Conclusions}
The z=3.65 quasar SDSS160705+533558 has been found to be bright in both the far-IR and submm. SED fitting indicates that the rest-frame far-IR luminosity is largely due to star formation with an inferred star formation rate of a few 10$^3 M_{\odot}yr^{-1}$. SMA imaging has found the submm emission to be extended in a linear north-south structure $\sim$ 36.2kpc long, dominated by a clump of size 16kpc to the north. The AGN is located roughly in the middle of the linear structure, but away from the submm peak. We conclude that an early stage merger between two galaxies, one of which is gas rich while the other hosts a supermassive black hole, is the most likely explanation for the unusual extended emission given the data currently available. Further observations of this unusual source, from optical to radio, are underway.

\acknowledgements
This work was funded in part by PPARC/STFC and NASA. The Submillimeter Array is a joint project between the Smithsonian Astrophysical Observatory and the Academia Sinica Institute of Astronomy and Astrophysics and is funded by the Smithsonian Institution and the Academia Sinica. The James Clerk Maxwell Telescope is operated by The Joint Astronomy Centre on behalf of the Science and Technology Facilities Council of the United Kingdom, the Netherlands Organisation for Scientific Research, and the National Research Council of Canada. Data were taken as part of JCMT project M05AU24. The authors thank Joe Silk, George Bendo and Pierre Chanial for useful discussions and the anonymous referee whose comments have significantly improved the paper.

\clearpage

\begin{deluxetable}{cc}
\tabletypesize{\scriptsize}
\tablecolumns{2}
\tablewidth{0pc}
\tablecaption{Basic Data for SDSS160705+533558 \label{tab1}}
\tablehead{
\colhead{Data}&\colhead{Value}
}
\startdata
RA                   &16 07 05.16\\
Dec                 &53 35 58.5\\
Redshift          &3.653\\
{\em u}        &23.61\\
{\em g}        &19.35\\
{\em r}         &18.15\\
{\em i}         &18.14\\
{\em z}        &17.98\\
J                    &0.36$\pm$0.04\\
H                   &0.31$\pm$0.06\\
K                   &0.34$\pm$0.07\\
3.6$\mu$m    &0.40\\
4.5$\mu$m    &0.40\\
5.8$\mu$m    &0.66$\pm$0.01\\
8.0 $\mu$m   &1.33$\pm$0.01\\
24 $\mu$m    &6.31$\pm$0.02\\
70  $\mu$m   &22.2$\pm$0.8\\
160 $\mu$m  &$<$104 (5$\sigma$)\\
450$\mu$m   &220$\pm$90 mJy\\
850$\mu$m   &31$\pm3$\\
1200$\mu$m &7.3$\pm$0.9\\
log$_{10}$(L$^{IR}_{total}$/L${\odot}$)&14.0---14.1\\
log$_{10}$(L$^{IR}_{SB} $/L${\odot}$)&13.3---13.7\\
log$_{10}$(L$^{IR}_{AGN}$/L${\odot}$)&13.8---13.9\\
SFR&3000-8000 M$_{\odot}/yr$\\
\enddata
\tablecomments{Fluxes, basic data and SED fit results for SDSS160705+533558.
SDSS fluxes are given in AB magnitudes all others are given as mJy. J,H and K fluxes are from 2MASS. Errors on IRAC1 and 2 fluxes are of order $5\%$. See Hatziminaoglou et al. (2005) for more details on SDSS and Spitzer fluxes and Perez-Fournon et al. (in preparation) for more details on submm fluxes. Derived luminosities are given for the restframe 1---1000$\mu$m waveband from our SED fitting.}
\end{deluxetable}

\clearpage

\begin{figure}
{\includegraphics[width=16cm,angle=0]{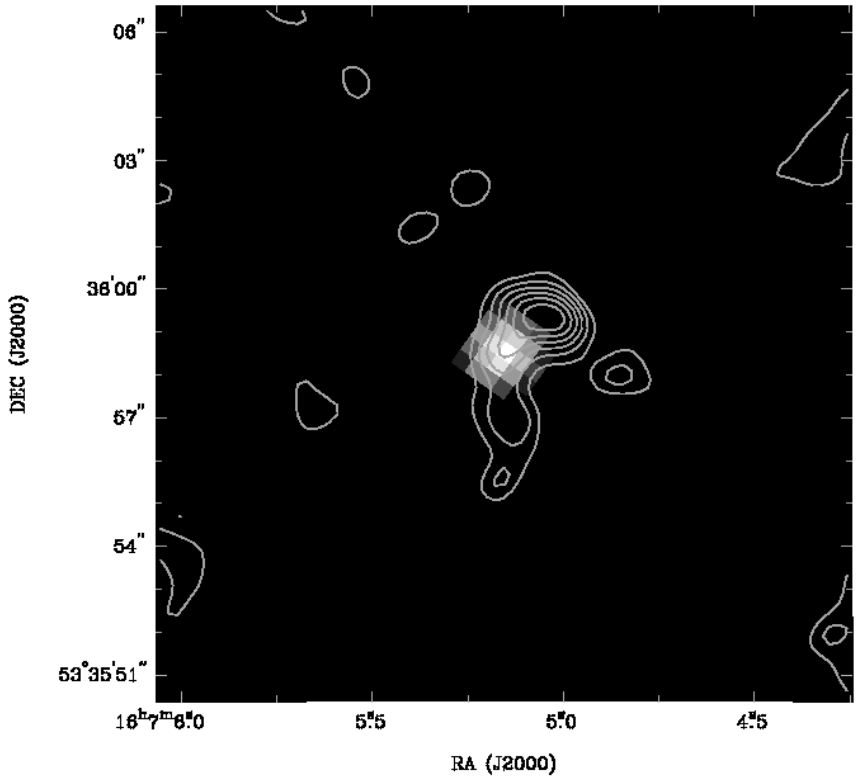}}
\caption{850$\mu$m Image of SDSS160705+533558}
Gray-scale optical image of SDSS160705+533558, overlaid with contours from the SMA submm image. Contours are in steps of 1$\sigma$ starting at 2$\sigma$, ie. at fluxes of 2.4, 3.6 etc. mJy for the 1.2mJy noise. Note the extended emission at submm wavelengths and the offset of the submm emission peak from the position of the optical quasar.
\end{figure}

\begin{figure}
{\includegraphics[width=12cm, angle=90]{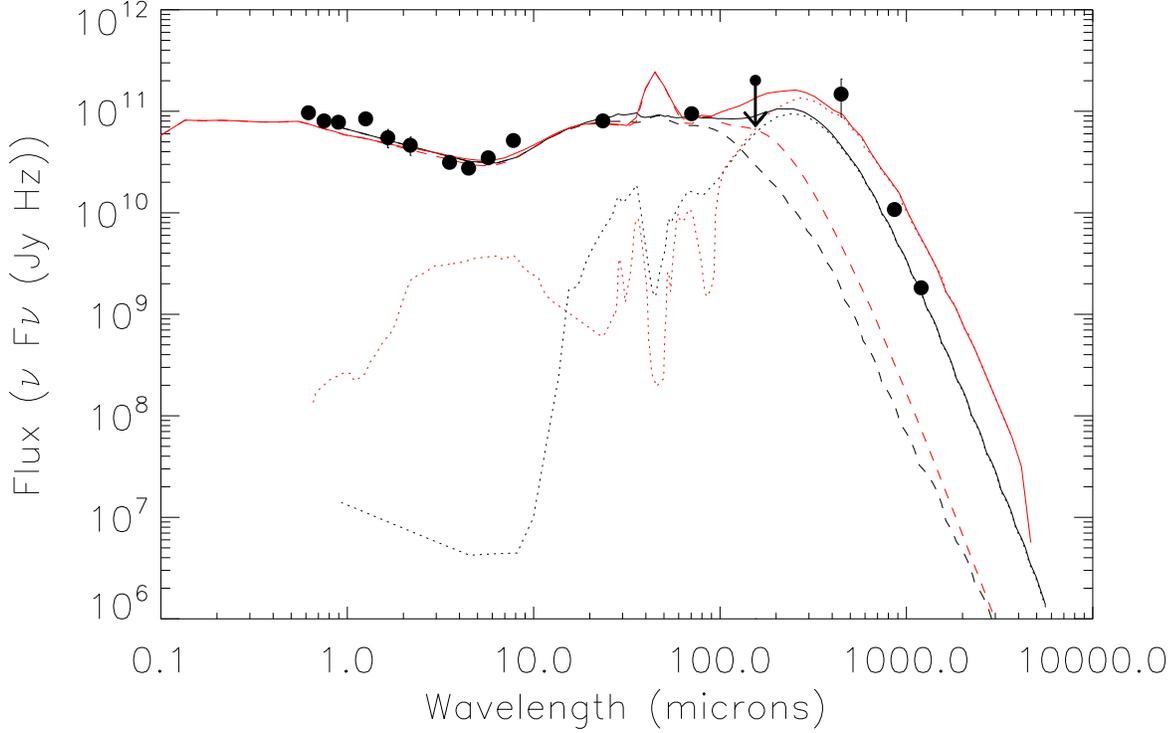}}
\caption{SED Fits for the IR-to-Submm fluxes of SDSS160705+533558}
We show the results for two different fits to the SED of SDSS160705+533558, one using the methods of Farrah et al. (2003), the other those of Hatziminaoglou et al. (2008). The Farrah fits are shown in black, the Hatziminaoglou fits are shown in red. For both fits we show the total SED as a solid line, starburst component as a dotted line and AGN as a dashed line.
The photometric data is shown as points, including the upper limit at 160$\mu$m. There are significant differences between the fits and neither is a perfect match to the data, but these two independent methods both reach the same conclusion that the mid-IR emission of this object is AGN dominated while the far-IR/submm emission, which we have resolved, is dominated by a starburst. 
\end{figure}

\end{document}